\documentclass[prc,reprint,onecolumn,amsmath,amssymb,superscriptaddress]{revtex4-1}
\usepackage{graphicx}
\newcommand{\be}{\begin{equation}}
\newcommand{\ee}{\end{equation}}
\newcommand{\beq}{\begin{eqnarray}}
\newcommand{\eeq}{\end{eqnarray}}
\newcommand{\ba}{\begin{array}}
\newcommand{\ea}{\end{array}}

\usepackage{color}

\begin{document}

\title{Comparative analysis of $\omega p$, $\phi p$, 
	and $J/\psi p$ scattering lengths from A2, CLAS, and 
	GlueX threshold measurements}

\date{\today}

%--------------------- AUTHOR LIST ---------------------------
\author{\mbox{Igor~Strakovsky}}
\altaffiliation{Corresponding author: \texttt{igor@gwu.edu}}
\affiliation{Institute for Nuclear Studies, Department of Physics, The 
	George Washington University, Washington, DC 20052, USA}

\author{\mbox{Lubomir~Pentchev}}
\affiliation{Thomas Jefferson National Accelerator Facility, Newport 
	News, Virginia 23606, USA}

\author{\mbox{Alexander~Titov}}
\affiliation{Bogoliubov Laboratory of Theoretical Physics,
	JINR, Dubna 141980, Russia}

\noaffiliation

%---------------------- ABSTRACT -----------------------------
\begin{abstract}
The high accuracy $\phi $-meson photoproduction data from the CLAS
experiment in Hall~B of Jefferson Laboratory allow us to determine
the near-threshold total cross section of the $\gamma p\to\phi p$
reaction and use it for evaluating the $\phi p$ scattering length
$\alpha_{\phi p}$. These data result in an absolute value of
$|\alpha_{\phi p}| = (0.063\pm 0.010)$~fm, which is smaller than
the typical hadron size. A comparative analysis of $\alpha_{\phi p}$
with the previously determined scattering lengths for $\omega p$ and
$J/\psi p$ from the A2 and GlueX experiments is performed.
\end{abstract}

\maketitle

%----------------------- Introduction ------------------------
%\section{Introduction}
%\label{sec:intro}

Since the discovery of the vector mesons ($\rho(770)$ in
1961~\cite{Erwin:1961ny}, $\omega(782)$ in 1961~\cite{Maglic:1961},
$K^\star(892)$ in 1961~\cite{Alston:1961nx}, $\phi(1020)$ in
1962~\cite{Connolly:1963pb,Schlein:1963zz}, $J/\psi(1S)$ in
1974~\cite{Aubert:1974js,Augustin:1974xw}, $D^\star$ in
1976~\cite{Goldhaber:1977qn,Nguyen:1977kk}, and $\Upsilon(1S)$ in
1977~\cite{Herb:1977ek}), they became attractive probes for the 
investigation of different aspects of the properties of
hadronic matter and hadronic interactions. In particular, exclusive
vector-meson photoproduction allows for the study of vector meson -
proton scattering and the evaluation of the 
corresponding scattering lengths $\alpha_{Vp}$, which may 
serve as a unique input for QCD-motivated models of vector 
meson-nucleon interactions. The absolute value of the scattering 
length may be determined from the near threshold vector-meson 
photoproduction total cross section by making use of the
vector-meson dominance (VMD) model~\cite{GellMann:1961tg}. 
VMD assumes that a real photon can fluctuate 
into a virtual vector meson, which subsequently scatters 
off the target proton~\cite{Titov:2007xb}. This method was 
used for the determination of the $\omega p$ and $J/\psi p$ 
scattering lengths~\cite{Strakovsky:2014wja,Strakovsky:2019bev}.

Recently, the CLAS Collaboration reported the first differential 
cross section measurements for the exclusive reaction $\gamma p\to\phi p$
near threshold~\cite{Dey:2014}. This is a unique experiment that measured
the differential cross sections from threshold at a photon energy of
$E_{\gamma} = 1.63$~GeV up to 2.82~GeV. The CLAS experiment used tagged
real photons produced from 4.023~GeV electrons by coherent Bremsstrahlung 
on a thin diamond radiator. The full acceptance of the 
detector in $\cos\theta$ span from -0.80 to 0.93, where 
$\theta$ is the $\phi$ meson center-of-mass (c.m.) production angle 
and is achieved by means of the CEBAF Large Acceptance 
Spectrometer (CLAS)~\cite{Mecking:2003zu} and the 
Bremsstrahlung photon tagging facility (``photon 
tagger")~\cite{Sober:2000we} in Hall~B of the Thomas Jefferson 
National Accelerator Facility (JLab). The $\phi$-meson was 
studied in both, the charged ($\phi\to K^+K^-$) and 
neutral ($\phi\to K^0_SK^0_L$, $K\bar{K}$) decay modes.

In this work, we report our determination of the total cross 
section $\sigma_t$ for the reaction $\gamma p\to\phi p$
near threshold using the recent differential cross sections 
measured by the CLAS Collaboration~\cite{Dey:2014},
and the estimation of the $\phi p$ scattering length 
$|\alpha_{\phi p}|$, which is compared with the previously 
evaluated scattering lengths for the $\omega p$ and $J/\psi p$ 
reaction~\cite{Strakovsky:2014wja,Strakovsky:2019bev}.

%------------ Total Cross section -----------------------
%\section{Total Cross section}
%\label{sec:sigma}

To determine the total cross sections from the CLAS differential 
cross-section $d\sigma/d\Omega(E_\gamma,\cos\theta)$, we use a 
series of Legendre polynomials $P_j(\cos\theta)$ (see, for 
instance, Ref.~\cite{Azimov:2016djk}):  
\begin{equation}
        d\sigma/d\Omega(E_\gamma,\cos\theta) 
        = \sum_{j=0}^{6}A_j(E_\gamma) P_j(\cos\theta)
        \label{eq:eq1}
\end{equation}
with integer $j$.
As expected for such a fit using orthogonal polynomials, the Legendre 
coefficients $A_j(E_\gamma)$ decrease markedly for large $j$. For the 
CLAS energies and precisions, a maximum value of $j = 6$ was found to 
be sufficient to describe the threshold data for $\cos\theta$ 
between -0.80 and 0.93. Recall that $\sigma_t = 4\pi A_0(E_\gamma)$. 
The best-fit results are summarized in Table~\ref{tbl:tab2}. Note 
that the large uncertainties at low energies result from the incomplete
angular coverage in $\cos \theta$.
%------------------------------------------
\begin{table}[htb!]

\centering \protect\caption{The total cross section $\sigma_t$ 
  for the reaction $\gamma p\to\phi p$ near threshold as a
  function of the beam energy $E_\gamma$. The errors 
  represent the total uncertainties resulting from the 
  quadrature sum of the statistical and systematic uncertainties of 
  the CLAS differential cross sections~\protect\cite{Dey:2014}. 
  The uncertainty of the beam energy is 12~MeV.}
\vspace{2mm}
{%
\begin{tabular}{|c|c||c|c|}
\hline
$E_\gamma$      & $\sigma_t$  & $E_\gamma$      & $\sigma_t$ \\
 (MeV)          & ($\mu$b)    &  (MeV)          & ($\mu$b)   \\
\hline
1673.1 & 0.1563 $\pm$ 0.0856 & 1959.9 & 0.3042 $\pm$ 0.0153 \\
1694.5 & 0.1460 $\pm$ 0.0686 & 1982.7 & 0.3467 $\pm$ 0.0141 \\
1716.1 & 0.1973 $\pm$ 0.0586 & 2005.6 & 0.3759 $\pm$ 0.0180 \\
1737.7 & 0.2126 $\pm$ 0.0386 & 2028.7 & 0.3434 $\pm$ 0.0152 \\
1759.4 & 0.2125 $\pm$ 0.0743 & 2051.8 & 0.3670 $\pm$ 0.0167 \\
1781.3 & 0.1862 $\pm$ 0.0709 & 2075.0 & 0.3485 $\pm$ 0.0146 \\
1803.2 & 0.2468 $\pm$ 0.0271 & 2098.4 & 0.3453 $\pm$ 0.0155 \\
1825.3 & 0.2237 $\pm$ 0.0166 & 2121.8 & 0.3578 $\pm$ 0.0165 \\
1847.5 & 0.2485 $\pm$ 0.0281 & 2145.4 & 0.3436 $\pm$ 0.0162 \\
1869.7 & 0.2856 $\pm$ 0.0223 & 2169.0 & 0.3367 $\pm$ 0.0164 \\
1892.1 & 0.2727 $\pm$ 0.0160 & 2192.8 & 0.3306 $\pm$ 0.0142 \\
1914.6 & 0.3140 $\pm$ 0.0197 & 2216.7 & 0.3266 $\pm$ 0.0133 \\
1937.2 & 0.3052 $\pm$ 0.0133 &        &                     \\
\hline
\end{tabular}} \label{tbl:tab2}
\end{table}
%------------------------------------------

%------------Scattering Length Determination -----------------------
%\section{Scattering Length Determination}
%\label{sec:SL}

As mentioned above, the near-threshold total cross sections 
of good accuracy allow for the extraction of the vector meson 
- proton scattering lengths as was done in 
Refs.~\cite{Strakovsky:2014wja,Strakovsky:2019bev}.  Below, 
we use the same approach for the extraction of the $\phi$-meson 
- proton scattering length.

In general, the total cross section of a binary reaction $ab\to
cd$ with particle masses $m_a + M_b < m_c + M_d$ can be written 
as $\sigma_t = \frac{q}{k} \cdot F(q,k,s)$, where $s$ is the 
square of the total c.m. energy, and $q$ and $k$ are the c.m.
momenta of the initial and final states, respectively. The 
factor $F(q,k,s)$ is proportional to the square of the invariant 
amplitude and does not vanish at threshold, where $E_\gamma\to 
E_{\rm thr}$, $q\to 0$, and $k$ is finite.  Thus, near 
threshold, $\sigma_t\to 0$ and is at least proportional to $q$.

\begin{figure}[htb]
\begin{center}
\includegraphics[height=4in, keepaspectratio, angle=90]{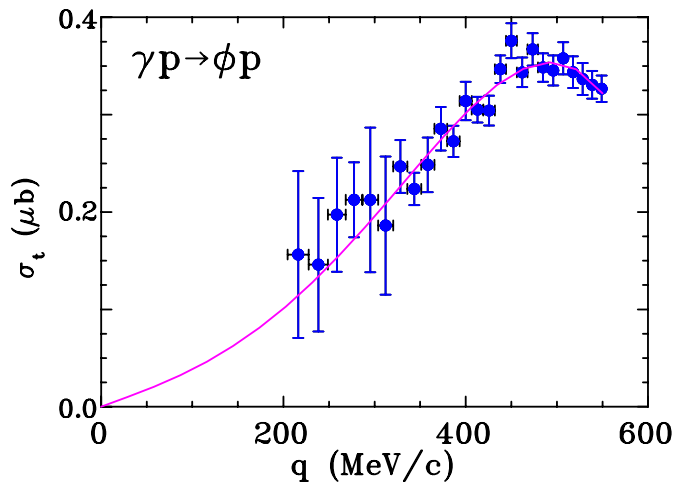}
\end{center}
\vspace{-7mm}
\caption{The total $\gamma p\to\phi p$ cross section $\sigma_t$ 
  (blue filled circles) derived from the CLAS data using 
  Eq.~(\protect\ref{eq:eq1}) (numerical results are available 
  in Table~\protect\ref{tbl:tab2}) is shown as a function of 
  the c.m. momentum $q$ of the final-state particles.  The 
  vertical error bars represent the total uncertainties of the 
  data summing statistical and systematic uncertainties in 
  quadrature, while the horizontal error bars reflect the energy 
  binning. The magenta solid curve shows the fit of the CLAS 
  data with Eq.~(\protect\ref{eq:eq2}).  Note that the first 
  data bin has a weighted average of $q$ = 216~MeV/c.} 
  \label{fig:fig1}
\end{figure}

Traditionally, the $\sigma_t$ behavior of a near-threshold 
binary inelastic reaction is described as a series of odd powers 
in $q$ (for details see Ref.~\cite{Strakovsky:2014wja}). In  the 
energy range under our study, we use:
\begin{equation}
        \sigma_t(q) = b_1q + b_3q^3 + b_5q^5,
        \label{eq:eq2}
\end{equation}
which assumes contributions from only the lowest $S$-, $P$-, 
and $D$-waves. Very close to threshold, the higher-order terms 
can be neglected and the linear term is determined by the 
$S$-wave only with a total spin of $1/2$ and/or $3/2$.  The 
fit of the total cross section using Eq.~(\ref{eq:eq2}) is 
shown in Fig.~\ref{fig:fig1} by the solid magenta curve. The 
best-fit results are summarized in Table~\ref{tbl:tab1}.
%------------------------------------------
\begin{table}[htb!]

\centering \protect\caption{The fit of the total cross section
  data using Eq.~(\protect\ref{eq:eq2}). The errors represent 
  the total uncertainties (summing statistical and systematic 
  uncertainties in quadrature).} 
\vspace{2mm}
{%
\begin{tabular}{|c|c|}
\hline
Parameter $b_i$            & Value     \\
\hline
$b_1$ [$\mu$b/(MeV/c)]     & (3.40$\pm$1.15)$\times$10$^{-4}$ \\
$b_3$ [$\mu$b/(MeV/c)$^3$] & (4.58$\pm$1.10)$\times$10$^{-9}$ \\
$b_5$ [$\mu$b/(MeV/c)$^5$] &(-12.48$\pm$2.53)$\times$10$^{-15}$ \\
\hline
$\chi^2$/d.o.f.               &  0.88          \\ \hline
\end{tabular}} \label{tbl:tab1}
\end{table}
%------------------------------------------
%-------------------------------------------------
\vspace{-12mm}
%-------------------------------------------------

For the evaluation of the absolute value of the vector meson - 
proton scattering length, we apply the commonly used and 
effective VMD approach, which links the near-threshold cross 
sections of the vector-meson photoproduction ($\gamma p\to V p$) 
and the elastic scattering ($V p\to V p$) processes via:
\begin{eqnarray}
	\frac{d\sigma^{\gamma p\to V p}}{d\Omega}|_{\rm thr}
	&=&\frac{q}{k} \cdot \frac{1}{64\pi}|T^{\gamma p\to Vp}|^2\nonumber\\
	&=&\frac{q}{k} \cdot \frac{\pi\alpha}{g_{V}^2}
        \frac{d\sigma^{V p\to V p}}{d\Omega}|_{\rm thr}
	=    \frac{q}{k} \cdot \frac{\pi\alpha}{g_{V}^2} 
	|\alpha_{Vp}|^2,
        \label{eq:eq77}
\end{eqnarray}
where $k$ is the photon c.m. momentum $k = (s - M_p^2)/2\sqrt{s}$,
$T^{\gamma p\to Vp}$ is the invariant amplitude of the 
vector-meson photoproduction, $\alpha$ is the fine-structure 
constant, and $g_V$ is the VMD coupling constant, related to the 
vector-meson electromagnetic (EM) decay width $\Gamma_{V\to e^+e^-}$
\begin{equation}
        g_{V} = \sqrt{\frac{\pi\alpha^2m_V}
        {3\Gamma_{V\to e^+e^-}}},
        \label{eq:eq78}
\end{equation}
where $m_V$ is the vector-meson mass.

Combining Eq.~(\ref{eq:eq2}) (which is also valid for
$\omega$- and $J/\psi$-photoproduction~\cite{Strakovsky:2014wja,
Strakovsky:2019bev}) and Eqs.~(\ref{eq:eq77}, \ref{eq:eq78}), 
one can express the absolute value of the scattering length 
as a product of the pure EM, VMD-motivated kinematic factor
$R_V^2 = {\alpha m_V k}/{12\pi\Gamma_{V\to e^+e^-}}$ and
the factor $h_{Vp} = \sqrt{b_1}$ that is determined by an 
interplay of strong (hadronic) and EM dynamics as
\begin{eqnarray}
	|\alpha_{Vp}| = R_V\,h_{Vp}.
	\label{eq:eq79}
\end{eqnarray}
In case of $\phi$-meson photoproduction, taking
$\Gamma_{\phi\to e^+e^-} = (1.27\pm 0.04)$~keV~\cite{Tanabashi:2018oca}
and $b_1$ from Table~\ref{tbl:tab1}, one gets $R_\phi = 
(343.0\pm 5.4)$~MeV$^{1/2}$
and $h_{\phi p} = (0.000184\pm 0.000032)$~fm/MeV$^{1/2}$,
which gives $|\alpha_{\phi p}| = (0.063\pm 0.010)~{\rm fm}$.

\begin{figure}[htb]
\begin{center}
\includegraphics[height=4in, keepaspectratio, angle=90]{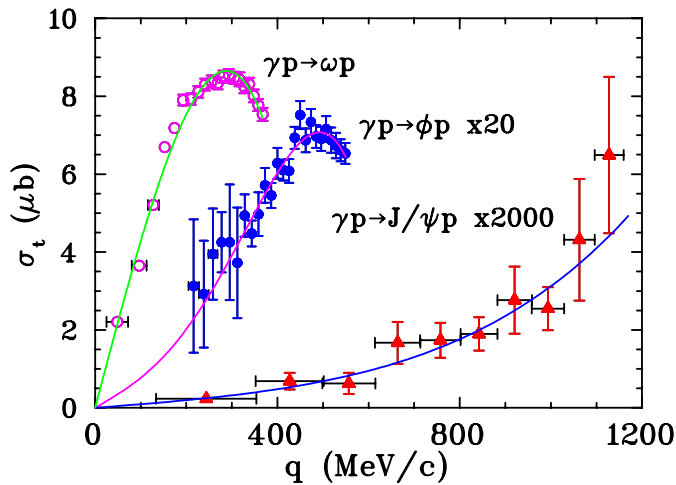}
\end{center}

\vspace{-5mm}
\caption{The total $\gamma p\to Vp$ cross section $\sigma_t$
  derived from the 
  A2 (magenta open circles)~\protect\cite{Strakovsky:2014wja},
  CLAS (blue filled circles)~\protect\cite{Dey:2014}, and
  GlueX (red filled triangles)~\protect\cite{Ali:2019lzf,
  Strakovsky:2019bev} data using Eq.~(\protect\ref{eq:eq1}) 
  is shown as a function of the c.m. momentum $q$ of the 
  final-state particles.  The vertical error bars represent 
  the total uncertainties of the data summing statistical 
  and systematic uncertainties in quadrature, while the 
  horizontal error bars reflect the energy binning. 
  Solid curves are the fit of the data with 
  Eq.~(\protect\ref{eq:eq2}).}  \label{fig:fig2}
\end{figure}

For the $\omega$ and $J/\psi$ mesons, Eq.~(\ref{eq:eq79}) 
results in $|\alpha_{\omega p}| = (0.820\pm 0.030)~{\rm fm}$,
and $|\alpha_{J/\psi p}| = (0.00308\pm 0.00055)~{\rm fm}$,
respectively~\cite{Strakovsky:2014wja,Strakovsky:2019bev}. 
The EM factors for the $\omega$ and $J/\psi$ mesons are 
close to each other, being 391~MeV$^{1/2}$ and 455~MeV$^{1/2}$, 
respectively.  Therefore, such a big difference in scattering 
lengths is determined mainly by the hadronic factor $h_{Vp}$, 
and reflects a strong weakening of the interaction in the 
$\bar cc-p$ system compared to that of the $\bar qq-p$ ($q = 
u, d$) configurations. The interaction in the $\bar ss-p$ 
configuration has an intermediate strength that is manifested 
in an intermediate value of the $\phi p$ scattering length.
Fig.~\ref{fig:fig2} illustrates the dramatic differences in 
the hadronic factors $h_{Vp} = \sqrt{b_1}$, as the slopes 
($b_1$) of the total cross sections at threshold as a function 
of $q$ vary significantly from $\omega $ to $J/\psi$.
%-------------------------------------------------
%-------------------------------------------------

The value for $|\alpha_{\phi p}|$ as determined in this 
paper from the CLAS data, is smaller than the results given
in the literature: 
$0.15$~fm from forward coherent $\phi$-meson 
photoproduction from deuterons near threshold by the LEPS
Collaboration~\cite{Chang:2007fc},
$(-0.15\pm 0.02)$~fm using a QCD sum rule analysis on the
spin-isospin averaged $\rho$, $\omega$, and $\phi$
meson-nucleon scattering~\cite{Koike:1996ga}, and
$\simeq$2.37~fm using the QCD van der Waals attractive 
$\phi N$ potential for the analysis of the $\phi$-nucleus 
bound states~\cite{Gao:2000az}. The latter value is more 
than an order of magnitude greater than the results using
experimental data and provides a problem for this particular 
potential model.
%-------------------------------------------------
\begin{figure}[htb]
\begin{center}
\includegraphics[height=3in, keepaspectratio, angle=90]{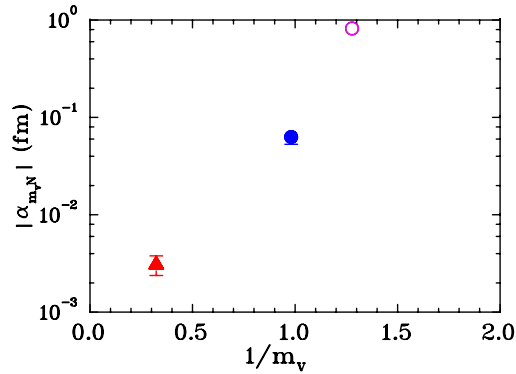}
\end{center}

\vspace{-5mm}
\caption{Comparison of the $|\alpha_{V p}|$ scattering 
  lengths estimated from vector-meson photoproduction 
  at threshold vs. the inverse mass of the vector meson.
  The magenta open circle shows the analysis of the A2 
  $\omega$-meson data~\protect\cite{Strakovsky:2014wja}, 
  the blue filled circle shows the current analysis of the 
  CLAS $\phi$-meson data~\protect\cite{Dey:2014}, and 
  the red filled triangle shows the analysis of the GlueX 
  $J/\psi$-meson data~\protect\cite{Ali:2019lzf,Strakovsky:2019bev}.}
  \label{fig:fig3}
\end{figure}
%-------------------------------------------------

Note that our value of $|\alpha_{\phi p}|$ is much smaller than 
the result from the A2 Collaboration at MAMI for the $\omega
p$ scattering length $|\alpha_{\omega p}|$ =
(0.82$\pm$0.03)~fm~\cite{Strakovsky:2014wja} and much larger
than the recent result from the GlueX data~\cite{Ali:2019lzf} for
the $J/\psi p$ scattering length $|\alpha_{J/\psi p}| = (0.00308\pm 
0.00055 ({\rm stat.}) \pm 0.00042 ({\rm 
syst.}))$~fm~\cite{Strakovsky:2019bev}. 
All results are shown in Fig.~\ref{fig:fig3} as a function of the 
inverse vector-meson mass.  Such a small value of the $|\alpha_{\phi p}|$
scattering length compared to the typical hadron size of 1~fm, indicates 
that the proton is more transparent for $\phi$-mesons compared to
$\omega$-mesons, and is much less transparent than for $J/\psi$ mesons.
Moreover, our analysis shows a non-linear exponential
increase $\alpha_{Vp}\propto\exp(1/m_V)$ with increasing $1/m_V$.

%--------------- Summary ----------------------------------
%\section{Summary}
%\label{sec:sum}

In summary, an experimental study of $\phi$-meson
photoproduction off the proton was performed by the CLAS
Collaboration at JLab~\cite{Dey:2014}. The quality of the CLAS
data near-threshold allows for the determination of the total 
cross sections of the reaction $\gamma p\to\phi p$ and for an 
estimation of the $\phi p$ scattering length within the VMD 
model. This results in an absolute value of the $\phi p$ 
scattering length that is smaller compared to the known 
theoretical prediction. We found 
$|\alpha_{J/\psi p}|\ll|\alpha_{\phi p}|\ll|\alpha_{\omega p}|$
and a strong exponential increase of $\alpha_{V p}$ with the 
inverse mass of the vector meson.

%-------------- Acknowledgments ----------------------------
%\section{Acknowledgments}

We thank Daniel Carman for valuable comments. This work 
was supported in part by the U.S. Department of Energy, Office 
of Science, Office of Nuclear Physics under Award No. 
DE--SC0016583 and Contract No. DE--AC05--06OR23177. 

%---------------------- REFERENCES ----------------------

%-----------------------------------------------
\end{document}